\documentclass[11pt,preprint]{article}
\usepackage[latin1]{inputenc}
\usepackage[T1]{fontenc}
\usepackage{amsmath}
\usepackage{amsfonts}
\usepackage[]{epsfig}
\usepackage{amssymb}
\usepackage{color}
\usepackage{graphicx}
\usepackage{tikz}
\usepackage{hyperref}
\usepackage{float}

  \newcommand{\be}{\begin{equation}}
\newcommand{\ee}{\end{equation}}

\newcommand{\ba}{\begin{eqnarray}}
\newcommand{\ea}{\end{eqnarray}}

\newcommand{\bea}{\begin{eqnarray}}
\newcommand{\eea}{\end{eqnarray}}

\newcommand{\fs}{i\kern+.01em\hbox{\raise.20ex\hbox{$/$}\kern-.58em$s$}}

\newcommand{\bs}{i\kern-.01em\hbox{\raise.25ex\hbox{$/$}\kern-.52em$b$}}
\newcommand{\qs}{/\kern-.52em s}

\newcommand{\dd}{\kern.06em\hbox{\raise.25ex\hbox{$/$}\kern-.40em$\partial$}}

\begin{document}
\tikzstyle{bag} = [text width=2em, text centered]
\tikzstyle{bag1} = [text width=5em, text centered]
\tikzstyle{end} = []
\title{$SU(2)$ Chern-Simons Theory Coupled to  Competing Scalars\\
}

\author{J.~M.~P\'erez Ipi\~na$^{a,b}$,
F.~A.~Schaposnik$^b$, G.~Tallarita$^c$
\\
~
\\
{\normalsize $^a\!$\it Departamento de F\'\i sica, Facultad de Ciencias Exactas y Naturales,}\\
{\normalsize \it Universidad de Buenos Aires, Ciudad Universitaria,}\\
{\normalsize \it  Buenos Aires 1428, Argentina}
\\ \vspace{-4 mm}\\
{\normalsize $^b\!$\it Departamento de F\'\i sica, Universidad
Nacional de La Plata}\\ {\normalsize\it Instituto de F\'\i sica La Plata-CONICET}\\
{\normalsize\it C.C. 67, 1900 La Plata,
Argentina}
\\ \vspace{-4 mm}\\
{\normalsize $^c\!$\it Departamento de Ciencias, Facultad de Artes Liberales,}\\ {\normalsize \it Universidad Adolfo Ib\'a\~nez, Santiago 7941169, Chile
 }
 }
\maketitle
\begin{abstract}
We study a spontaneously broken $SU(2)$ Chern-Simons-Higgs mo\-del   coupled though a Higgs portal to an uncharged  triplet scalar with a vacuum state  competing with the Higgs one.  We find vortex-like solutions  to the field equations in different  parameter space regions. Depending on the scalar coupling constants  we find a parameter  region in which the competing order creates a halo about the Chern-Simons-Higgs vortex core, together with two other  regions, one where  no vortex solutions exist, the other where ordinary Chern-Simons-Higgs vortices can be found. We derive the low-energy theory for the moduli fields on the vortex world sheet and also discuss the connection of our results with those found in studies of competing orders in high temperature superconductors.
\end{abstract}
\section{Introduction}
Among the many applications of vortex-like solutions in
 gauge theories  there has been renewed interest in the study of models in which there are two scalars with different ground states, a subject which is relevant both in high energy and condensed matter physics.

 In high energy physics, the existence of a charged scalar with a nontrivial vacuum expectation value in the core of a vortex can lead to striking effects originally signaled in \cite{Witten} in the context of superconducting strings. Also (center) vortices are relevant to the study of confinement (see \cite{tH}- \cite{Greensite} and references therein for more recent results)  and in  the effective descriptions in the case of $2+1$ space-time where the Chern-Simons term  plays a central role \cite{PRD}.
 There have also been recent and interesting applications  in the construction  of  low-energy Lagrangians in the world-sheet of the Abrikosov-Nielsen-Olesen string by the addition of an extra non-Abelian moduli field  \cite{Shifman}-\cite{Shifman2}.

 Concerning condensed matter, the presence of the two scalars
 allows for the existence of  competing vacuum states  in systems in which the superconductivity order is suppressed in the vortex core where a "stripe order" formation takes place \cite{Zhang}-\cite{Arovas}, \cite{Sachdev}. These ideas are at present the basis of many investigations on phase transitions, as in particular those referred to competing order in mixed states of high-temperature superconductors \cite{Ed1}-\cite{Ed2}.

Gauge theories with Maxwell and  Chern-Simons terms coupled to scalars in $d=3$ space-time dimensions have vortex solutions, both for Abelian \cite{PK} and also for non-Abelian gauge groups \cite{deVS} provided one includes as many scalars as required to have maximum symmetry breaking. In particular, in the case of an $SU(N)$ gauge group one needs   $N$ scalars in the adjoint representation in order to have topologically non-trivial (vortex) solutions of the field equations. In that case,  the surviving symmetry is $\mathbf{Z}_N$ and the vortex solutions belong to $(N-1)$ homotopy classes.
In the particular case of pure Chern-Simons-Higgs theories,  first order field equations exist and a Bogomolny bound for vortex solutions both in Abelian \cite{Ko}-\cite{JW} and non-Abelian \cite{CLS} cases were found.

In this work we shall consider an $SU(2)$ Chern-Simons gauge theory coupled to two scalar triplets $\vec\phi,\vec\psi$ with an appropriate potential leading to a symmetry breaking ground state. This model will be coupled  to an  $O(3)$ triplet $\vec \chi$, uncharged under the gauge group, with a potential   $U(|\vec \phi|, |\vec\chi|)$ such that in certain regions of the parameter space the theory exhibits  a competing non trivial $\chi$ vacuum state and vortex solutions. We shall also discuss  the moduli fields associated to the  $\vec \chi$ and the resulting low energy dynamics of the vortices.

The paper is organized as follows. In section 2 we introduce the  pure Chern-Simons-Higgs (CSH)  model, discuss the pattern of symmetry breaking  and
the topology of  CSH vortex configurations. We present  in section 3 the action governing the dynamics of the competing  triplet   $\chi$ field coupled to the CSH model via a
  Higgs portal mixing   and  analyze the conditions leading to  non-trivial vacuum states for $\phi$ and $\chi$.  We briefly discuss
  the non-Abelian moduli   fields associated to $\chi$  localized on the vortex  world sheet in  section 4  and then present in section 5 the solutions
to the coupled field equations discussing in detail the vortex solutions in different  parameter space regions. Finally, in
section 6 we summarize and discuss  our results, in particular stressing its connection with superconductivity with compeeting orders.

\section{Non-Abelian Chern-Simons vortices}
 The existence of classic  non-trivial vortex solutions in non-Abelian gauge theories coupled to scalars requires maximum symmetry breaking \cite{deVS}. Otherwise, no topologically stable non-trivial solutions can be found, although quantum corrections can stabilize them, for example if  heavy  fermions are coupled to the Abelian Higgs model \cite{WQG}.

 To this end, in the case of an $SU(2)$ Chern-Simons gauge theory one has  to couple the gauge field $A_\mu$ to two scalars in the adjoint representation, $\vec \phi,\vec\psi$.  We shall briefly outline in this section the construction of such topologically nontrivial configurations as derived in \cite{CLS}.

 Dynamics of the model is governed  by the following $2+1$ dimensional action
\begin{equation}
\begin{aligned}
S_{CSH} =
&    \int d^3x \left(\vphantom{\int_A^B} \dfrac{\kappa}{4}  \varepsilon^{\mu \nu \gamma} \left(\vphantom{\frac12}  \vec{F}_{\mu \nu} \cdot \vec{A}_\gamma  - \dfrac{e}{3} \vec{A}_\mu \cdot ( \vec{A}_\nu  \wedge \vec{A}_\gamma) \right)\right.\\
&\left. + \dfrac{1}{2} (D_\mu \vec{\phi})\cdot(D^\mu \vec{\phi}) + \dfrac{1}{2} (D_\mu \vec{\psi})\cdot(D^\mu \vec{\psi}) - V(\vec{\phi}, \vec{\psi}) \right)
\vphantom{\int_A^B}
\end{aligned}
\label{LCSH}
\end{equation}
Here the gauge field $A_\mu$ ($\mu = 0,1,2$) takes value in the algebra of $SU(2)$,
\begin{equation}
A_\mu = A_\mu^a T^a \equiv \vec{A_\mu} \cdot \vec{T}
\end{equation}
with generators $T^a$ ($a=1,2,3$) satisfying
\begin{equation}
[T^a,T^b] = i \varepsilon^{a b c} T^c , \ \
\lbrace T^a,T^b \rbrace  = \dfrac{1}{2} \delta^{ab} , \ \
Tr(T^aT^b)= \dfrac{1}{2} \delta^{ab}
\end{equation}

With this, the field strength $\vec{F}_{\mu \nu} $ and covariant derivatives can be written as
\bea
\vec{F}_{\mu \nu} &=& \partial_\mu \vec{A_\nu} - \partial_\nu \vec{A_\mu} + e \vec{A_\mu} \wedge \vec{A_\nu} \\
D_\mu\vec\phi &=&  \partial_\mu \vec{\phi} + e \vec{A_\mu} \wedge \vec{\phi}
\eea
and an analogous formula for $D_\mu\vec{\psi}$.

Concerning the symmetry breaking potential it takes the form
\begin{equation} \label{Vgeneral}
V(\vec{\phi}, \vec{\psi}) = V_1(\vec{\phi}) + V_2(\vec{\psi}) + g''(\vec{\phi} \cdot \vec{\psi})^2
\end{equation}
where $V_1$ and $V_2$ have to be chosen so that  asymptotically gauge symmetry is maximally broken:
\begin{equation} \label{vacios}
\begin{aligned}
&\lim_{\rho \to \infty} |\vec{\phi}| = \phi_0 \\
&\lim_{\rho \to \infty} |\vec{\psi}| = \psi_0 \\
&\lim_{\rho \to \infty} \vec{\phi}\cdot\vec{\psi}= 0
\end{aligned}
\end{equation}
Note that the third term in Eq.\eqref{Vgeneral} is the one that forces orthogonality between $\vec\phi$ and $\vec\psi$  thus ensuring maximal symmetry breaking and the existence of topologically non-trivial vortex solutions. As a result,  the invariant group of the vacuum is $\mathbf{Z}_2$  and the relevant homotopy group is $\Pi_1(SU(2)/\mathbf{Z}_2) = \mathbf{Z}_2$.

Concerning the symmetry breaking potentials    we shall choose for $V_1(|\vec \phi|)$ the sixth order one in order to compare our results whith those of a pure CS-Higgs theory at the Bogomolnyi point \cite{CLS},
\be
V_1(|\vec \phi|) = \frac{\lambda^2}2 |\vec \phi|^2 (|\vec \phi|^2 - \phi_0^2)^2
\label{8}
\ee
As we shall see below, in view of the minimal energy ansatz that we shall make it will not be necessary to explicitly choose $V_2(|\vec \psi|)$.

The topological charge associated to the non-Abelian vortex configurations can be calculated  as usual via a Wilson loop,
\be
Q_T= \frac12 {\rm Tr} {\cal P} \exp\left( ie\oint_{C_\infty} \!\!\!A_\mu dx^\mu\right)
\ee
with Tr the $SU(2)$ trace, ${\cal P}$ the path-ordering operator  and $C_{\infty}$ a closed curve at infinity. We shall explicitly see that the suitable ansatz leads to $Q_T = (-1)^n$ so that  that there are two topologically
inequivalent configurations, the ``trivial'' one with   $Q_T = 1$
($n =2l, l\in \mathbf{Z}$) and those with $Q = -1$   the ``non-trivial'' one
($n = 2l + 1$).

 It should be stressed that, as  already signaled in the pioneering work of Deser, Jackiw and Templeton \cite{DJT},  the non Abelian Chern-Simons action $S_{CS}[A]$ is not invariant under large gauge transformations $g \in SU(2)$
\begin{equation}
S_{CS} [A^g] = S_{CS}[A] +  {8 \pi^2}  \kappa \omega[g]
\end{equation}
with $w[g]$ the winding number associated to the gauge transformation,
\begin{equation}
\omega[g] = \frac{1}{24 \pi^2}\text{Tr} \epsilon^{\mu \nu \gamma} \int g^{-1} \partial_\mu g g^{-1} \partial_\nu g g^{-1} \partial_\gamma g = n  \in \mathbf{Z}
\end{equation}
Then, in order to have a gauge invariant partition function, $\kappa$ should be chosen as
 \begin{equation}
\kappa = \frac{ k }{4 \pi} \ , \ k \in \mathbf{Z}
\end{equation}

As discussed in ref. \cite{CLS}  an axially symmetric ansatz allows to find vortex solutions with $Q_T=-1$ to the field equations arising from action \eqref{LCSH}. Moreover, first order self-dual  equations  exist for a choice of symmetry breaking potential as defined in eq. \eqref{8}  for one of the two scalars, say $\vec \phi$. The lowest energy vortex  solution will then correspond  to the case in which the second scalar ($\vec\psi$) is chosen everywhere in its vacuum expectation value so that its role is just to achieve complete symmetry breaking \cite{deVSreply}.

Because of the nature of the Gauss law in Chern-Simons theories vortex solutions should have not only magnetic flux but also electric charge and angular momentum \cite{deVS}, all of them quantized. Finally, it should be noticed  that in the case of non-Abelian gauge theories both for Yang-Mills and CS cases, the bound of the energy is not of topological character. Indeed, as shown in  \cite{CLS}  for the non-Abelian Chern-Simons case,   the  Bogomolny bound for the energy in $2+1$ dimensions is  not just  proportional  to the vortex topological charge $Q_T$ but  to the sum of $(Q_T + 2k)$. Here $k \in Z$ is an integer related to the Cartan subgroup of the gauge group and depends  on the gauge element one chooses to use in the asymptotic gauge field behavior. The same happens for the energy per unit length in $3+1$ dimensional Yang-Mills theory as first shown in \cite{CLSotro}.

\section{Adding a competing scalar}

Let us now couple action $S_{CSH}[A,\phi,\psi]$ with an
$O(3)$ global action  $S_\chi$ for a triplet scalar $\chi$ coupled to the Higgs field $\phi$ in such a way that     its vacuum is in competition with the one of the dynamical scalar $\phi$.

\begin{equation}
S_\chi = \int d^3x \left(\frac{1}{2} \partial_\mu \vec\chi  \partial^\mu \vec\chi - U(\vec\chi,\vec\phi)\right)
\label{13}
\end{equation}
with the potential \cite{Shifman}
\begin{equation} \label{potU}
U(\vec \chi,\vec \phi) = \gamma \left[ (-\mu^2 + |\vec{\phi}|^2) |\vec{\chi}|^2 + \beta (|\vec{\chi}|^2)^2\right]
\end{equation}

Before a detailed analysis of the potential $U$, let us stress that when one couples action \eqref{13}  to the Chern-Simons-Higgs  action  \eqref{LCSH}  with its potential $V(|\vec\phi|)$  chosen so as to have BPS equations, the resulting action cannot have first order (selfdual) BPS equations. A simple way to see this is to recall that selfduality is intimately related to the possibility of extending the original bosonic action to a supersymmetric one \cite{WO}. Now the supersymmetric extension of  Higgs-portal term $|\vec{\phi}|^2 |\vec{\chi}|^2$  in  \eqref{13} requires the introduction of a second gauge field coupled to $\chi$ and the addition of a gauge mixing coupling \cite{AIST}-\cite{IST} which would then completely change the character of the model.

Potential parameters,  $\gamma$ and $\mu$ are real and positive constants. We are interested in finding vortex like solutions in which there is a competition between the vacuum states of scalars $\phi$ and $\chi$. Vorticity implies that $|\phi| \to \phi_0$ at infinity and it should vanish at the origin,   $\phi(0)=0$. Concerning $\chi$, we shall impose that $\lim |\chi| \to 0 $ asymptotically and that it takes a non-zero value at the origin
 fixing   $|\vec\chi |$ but not its direction so that at short distances the $O(3)$ symmetry  is spontaneously broken to  $O(2)$.
Moreover we want  the length scale of variation of the $\phi$ field to be larger than or of the order of the    $\chi$ length scale, this implying
\begin{equation} \label{escalaescalares}
\sqrt{\gamma \mu^2}   \gtrsim \lambda^2 \phi_0^2
\end{equation}

This condition is usually required in the case of Maxwell gauge dynamics \cite{Shifman}-\cite{Shifman2} so that $\chi$ is localized in the center of the vortex core, where the magnetic field has its maximum.  In the present  case the Chern-Simons gauge dynamics defined in $2+1$ dimensions forces the magnetic field to vanish at the origin (see  refs.~\cite{Ko}-\cite{JW}) so in the spatial  plane the magnetic field is not concentrated in the vortex core (a disc) as with Maxwell theory,  but in the annulus centered at the origin, bounded by  concentric circles of radii $R_{min}, R_{max}$. This implies that  there are two regions where the competing field $\chi$ can form a ``halo'': inside the smaller radius circle of the annulus or outside the larger one.  We shall discuss in detail in section 5 the reasons behind condition \eqref{escalaescalares} and the nature of the observed halo.

 The action $S$ for the model with two competing scalars that we will investigate is then  defined as
 \be
 S = S_{CSH}[A,\phi,\psi] + S_\chi[\chi,\phi]
 \label{completa}
 \ee
Dimensions of fields and parameters are: $[A_\mu]= m; [\phi]=[\psi]= [\chi]=m^{1/2}$; $ [\mu] =m^{1/2};  [\gamma] = m;  [\kappa] = [e] = [\beta] = [\lambda] = 0$.

Since we choose the field $\vec \psi$ to be constant everywhere, the field equations reduce to
\begin{align}
&D_\mu D^\mu \vec{\phi} = -\frac{\partial V_1(\vec{\phi})}{\partial \vec{\phi}} - \frac{\partial U(\vec{\chi},\vec{\phi})}{\partial \vec{\phi}} \label{eomgen1}\\
&\partial_\mu \partial^\mu \vec{\chi} = -\frac{\partial U(\vec{\chi},\vec{\phi})}{\partial \vec{\chi}} \label{eomgen2}\\
&\frac{\kappa}{2} \epsilon^{\mu \nu \gamma}\vec{F}_{\nu \gamma} = \vec{j}^\mu \label{eomgen3}
\end{align}
with
\begin{equation}
\vec{j}^\mu = e (D^\mu \vec{\phi}) \wedge \vec{\phi}
\end{equation}
The scalar and gauge excitation masses are:
\be
 m_\phi = 2 \lambda \phi_0^2 \,, \;\;\;\;  m_A=\dfrac{e^2 \phi_0^2}{\kappa}
\ee
Now, using the Gauss law to eliminate $\vec A_0$ in terms of the magnetic field,
\be
(\vec A_0 \wedge \vec \phi)\wedge \vec\phi = \frac{\kappa}{e^2}  {\vec B}
\ee
we ge for the energy
\begin{equation}
\mathcal{E}_{CSH\chi} = \int d^2x \left[ \frac{\kappa}{2 e^2 |\vec{\phi}|^2} |\vec{B}|^2 + \frac{1}{2}|D_i \vec{\phi}|^2 + V_1(\vec{\phi}) + \frac{1}{2} |\partial_i \vec{\chi}|^2 + U(\vec{\chi},\vec{\phi}) \right]
\label{inserting}
\end{equation}

In order to find vortex-like solutions we propose
 the following static axially symmetric ansatz
\begin{eqnarray} \label{ansatz}
\vec{A}_\varphi &=& - \dfrac{1}{e} \dfrac{a(\rho)}{\rho}\begin{pmatrix} 0\\0\\1 \end{pmatrix}\,, \;\;\;\;   \vec{A}_0  =  - \dfrac{1}{e} a_0(\rho)\begin{pmatrix} 0\\0\\1 \end{pmatrix}
\nonumber\\~\\\nonumber
\vec{\phi} &=& \phi_0\ f(\rho)\begin{pmatrix} \cos\varphi\\\sin\varphi\\0 \end{pmatrix}\,, \;\;\;\;   {\vec{\chi} = \chi_0\ h(\rho)\begin{pmatrix} 0 \\ 0 \\ 1 \end{pmatrix}}
\end{eqnarray}
with $\chi_0=\mu/\sqrt{2\beta}$, which corresponds to the minimum of the $U$ potential at the origin. Complete $SU(2)$ symmetry breaking (except for the $SU(2)$ center) implies that
\be
 \vec{\psi} = \psi_0\begin{pmatrix} 0\\0\\1\end{pmatrix} \label{ansatzchi}
 \ee
with $\psi_0$ the minimum of the $V_2(|\psi|)$ potential.
Finite energy solutions imply  the following conditions
\[
a(\infty) = f(\infty) = 1,\ \  a_0(\infty) = h(\infty)=0 \]
\be a(0) = f(0) = 0,\ \ a_0(0) = b 
\label{an}
\ee
 Here $b$ is a constant that is fixed through the Gauss law in terms of $\kappa, e$ and the ratio of the magnetic and Higgs field squared modulus at the origin.  Concerning the $\chi$ field we shall impose $h'(0)=0$. One can see this by making a Frobenius analysis of the field equations close to the origin. All odd powers of $\rho$ in the $h$ expansion vanish, which in turn implies the derivative there must also vanish (see below).

Inserting the  vortex ansatz in eq. \eqref{inserting} we get for the vortex energy
\begin{align}
&\mathcal{E} = 2 \pi \phi_0^2 \int \xi d\xi \left[ \frac{a'(\xi)^2}{2 \xi^2 f^2(\xi)} + \frac{1}{2} \left( f'(\xi)^2 + \frac{f^2(\xi)}{\xi^2}(a(\xi)-1)^2 \right) + \right. \nonumber \\
& \left. + \frac{c_1}{8} f^2(\xi)(f^2(\xi)-1)^2 + \frac{c_2}{2 c_3} h'(\xi)^2  +  c_3 h^2(\xi)  \left( \frac{1}{2} \left(\frac{\mu}{\phi_0}\right)^2 (h^2(\xi)-2) + f^2(\xi) \right) \right]  \nonumber \\
& \ \ \ \equiv \int \epsilon(\xi) d\xi
\end{align}
    where we have introduced the dimensionless radial variable $\xi = m_A \rho$,  $f' = {d}/{d\xi}$,
   and the dimensionless parameters
\begin{equation}
c_1 = \dfrac{m_\phi^2}{m_A^2} ,\;\;\;\; c_2 = \dfrac{\gamma \chi_0^2}{m_A^2} , \;\;\;\;\ c_3 = \dfrac{\gamma \phi_0^2}{m_A^2}
\end{equation}
Finally,
 $\epsilon (\xi)$ is the energy density.

Concerning the field equations, after inserting  the ansatz \eqref{ansatz}-\eqref{ansatzchi} in eqs. \eqref{eomgen1}-\eqref{eomgen3} we get the following system of non linear coupled equations
\begin{align}
&f''(\xi)+\xi^{-1}f'(\xi)-\xi^{-2}f(\xi)(1-a(\xi))^2 + (a'(\xi))^2/(\xi^2 f^3(\xi)) = \nonumber \label{eomf} \\
&=\frac{c_1}{4} f(\xi)(f^2(\xi)-1)(3f^2(\xi)-1)+ 2 c_2 f(\xi) h^2(\xi)\\
&h''(\xi)+\xi^{-1}h'(\xi) = 2 \beta c_2 h(\xi) \left( h^2(\xi) -1\right) + 2 c_3 h(\xi) f^2(\xi) \label{eomh}\\
&a''(\xi) - \left[ \frac{1}{\xi} + 2 \frac{f'(\xi)}{f(\xi)}\right]a'(\xi) + f^4(\xi) [1-a(\xi)] = 0 \label{eoma}
\end{align}
Here we have used the Gauss Law to write $\vec A_0$ in terms of the magnetic and Higgs fields.

As mentioned above, following a Frobenius analysis of the field equations close to the origin, the Neumann condition imposed on $\chi$ can be justified. We set $a(0)=a'(0)=0$ in eqs. \eqref{eomf}-\eqref{eomh}  and we expand  the scalar fields $f(\xi)$ in odd powers of $\xi$ and the $h(\xi)$ field in all powers of $\xi$:
\begin{eqnarray} \nonumber
f &=& \xi f_1 + \xi^3 f_2 + \ldots  \\
h &=& h_0 + \xi h_1 + \xi^2 h_2 + \ldots \nonumber
\end{eqnarray}
where $\ldots$ denote higher powers in $\xi$.
Then, inserting the expansions in the field equations, a $\xi$ power by power analysis of the equations reveals that all odd powers of $\xi $ in the $\chi$ expansion vanish, which in turn implies its derivatives must vanish.

Let us end this section stresing that due to the lack of a second gauge field with a suitable   coupling to $\vec A_\mu$ and to $\vec \chi$ one cannot reduce the second order field equations to first order BPS ones.

\section{The $\chi$ field orientational moduli}

Before finding the axially symmetric solutions to the field equations we shall discuss the orientational moduli associated to the presence of the $\chi$-field \cite{Shifman}. To this end, instead of considering the  specific choice proposed for $\chi$ in ansatz (\ref{ansatz}),  we allow for a more general  parametrization which implies a time dependence,
\be
\chi^i =\sqrt{\frac{\mu^2}{2\beta}} h(\rho)n^i(t)
\label{g}
\ee
where $n^i(t)$ is  satisfying
\be
n^i(t) n^i(t) = 1
\ee
Inserting \eqref{g} in the action \eqref{completa} we obtain  the low-energy action for the orientational moduli
\be
S_{\cal O} = \frac{I_{\cal O}}2   \int dt\; n^i(t) n^i(t)
\label{gi}
\ee
where
\be
I_{\cal O} =  \frac{4\pi\mu^2}{\beta} \int d\rho \rho^2 h(\rho)^2.
\ee

This is the action for the $CP(1)$ non-linear sigma model, as expected from the pattern of global symmetry breaking in the $\chi$ sector: $SU(2)/U(1) \rightarrow CP(1)$. We can proceed to quantize it by parametrizing the unit vector $n^i$ in terms of polar and azimutal angles $(\alpha,\vartheta)$,
\be
\vec n = \left(
\begin{array}{l}
\cos\alpha\sin\vartheta\\
\sin \alpha\sin\vartheta\\
\cos\vartheta
\end{array}
\right)
\ee
Action \eqref{gi} is the action for a symmetric top with moment of inertia $I_{\cal O}$ so that upon canonical quantization the Hamiltonian operator reads
\be
\hat H = -\frac1{2I_{\cal O}} \left(
\frac1{\sin\vartheta}\frac{\partial}{\partial\vartheta}\left(\sin\vartheta \frac{\partial}{\partial\vartheta}   \right)
+ \frac1{\sin^2\vartheta  }\frac{\partial^2}{\partial^2\alpha}
\right)
\ee
which of course has spherical harmonics eigenfunctions with eigenvalues
\be
E_{\mathcal L} = \frac1{2I_{\cal O}} {\mathcal S}({\mathcal S} + 1),
\ee
where $S$ is an integer. Therefore, the vortices found below for which $\chi\neq0$ correspond to vortices with "isospin".  Note that these orientational gapless excitations can be lifted by the introduction of a "spin-orbit" coupling of the form $\epsilon(\partial_i \chi^i)^2$ to the Lagrangian \cite{Shifman:2013oia}. As shown in \cite{Shifman2}, this is the starting point for an interesting correspondence between this system with a new kind of superconducting liquid crystal phase, with the cholesteric formed by the introduction of a $\eta\epsilon_{ijk}\chi_i\partial_j \chi_k$ term to the original Lagrangian. We will leave the detailed analysis of this connection (with regards to this specific model) to a future publication.

\section{Solutions}
Action \eqref{completa} has the trivial extremum $\vec \phi=\vec\chi=0$. Of course, for $\chi =0$ one has the  non-Abelian vortex CS-Higgs solution discussed in \cite{CLS}. There exists also a spatially uniform solution in which both   $\phi$ and  $\chi$ are constant (with $A_\mu = 0$),
\bea
|\vec \phi| &=&  \phi_0 \nonumber\\
|\vec \chi| &=& \sqrt{\frac{\mu^2 - \phi_0^2}{2\beta}}
\label{solconstante}
\eea
Note that this solution is only valid provided $\phi_0^2 / \mu^2  < 1$ and was also found in  \cite{Ed1} where it corresponds to the region of coexisting order, in the schematic zero-temperature diagram that the authors present.

Concerning $(\vec A,\vec \phi)$-vortex configurations and fully non-trivial $\chi$, we have found, numerically, solutions of eqs.\,\eqref{eomf}-\eqref{eoma} using a second order central finite difference procedure with accuracy ${\cal O}(10^{-4})$. We tested the solver for the particular case $\chi=0$ and we accurately reproduced the exact result of the Bogomolnyi lowest  bound for the lowest energy  $n=1$ vortex,  $E=\phi_0^2 \pi$ \cite{CLS}.

\begin{figure}[h]
\centering
\includegraphics[scale=.5]{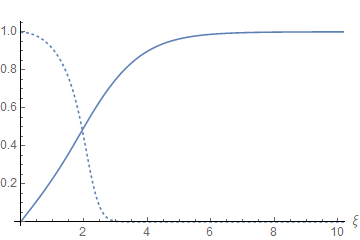}
\caption{Higgs (solid line) and $\chi$ (dashed line) profiles ($f(\xi),h(\xi)$) for $e=2$, $\lambda=1$, $\kappa=2$ ($c_1=1$), $\beta=160$, $\gamma=200$, $\mu / \phi_0=0.8$ (dimensionful parameters have been normalized by the gauge field mass).}
\label{figscalars:c1=1}
\end{figure}
\begin{figure}[h]
\centering
\includegraphics[scale=.5]{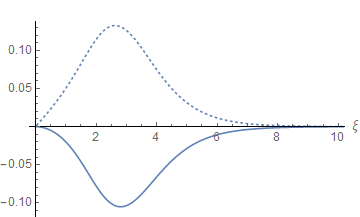}
\caption{Magnetic field (solid line) and electric field (dashed line) for $e=2$, $\lambda=1$, $\kappa=2$ ($c_1=1$), $\beta=160$, $\gamma=200$, $\mu / \phi_0=0.8$ (dimensionful parameters have been normalized by the gauge field mass).}
\label{figelectro:c1=1}
\end{figure}

Typical field profiles of the scalar and electromagnetic fields are shown in Figures \ref{figscalars:c1=1}-\ref{figelectro:c1=1} for the case in which the gauge coupling constant $e$, the Chern-Simons coefficient $\kappa$ and the Higgs potential coupling constant $\lambda$ are chosen to the values in which the CSH model is defined at the Bogomolnyi point ($m_\phi = m_A$).

The competing orders are clearly shown in the $\vec \phi, \vec \chi$ profiles in Fig. \ref{figscalars:c1=1} where one can see that the coherence lengths of both scalar fields are of the same order. This is a relevant property for condensed matter models that describe the quantum phase transition between a pure superconducting phase from one in which a competing order coexists with superconductivity. The result is a
``halo''  about the vortex core whose existence has  been inferred from the charge-stripe order that has been found experimentally (see \cite{Ed1} and references therein).

 The electromagnetic field profiles  in Fig. \ref{figelectro:c1=1} are qualitatively similar to those in the pure CSH model. As it happens when a CS term is present, in addition to the magnetic filed, vortices are electrically charged and both fields should vanish at the origin where the Higgs field has to vanish to ensure regularity of the solution.

\begin{figure}[h]
\centering
\includegraphics[scale=.5]{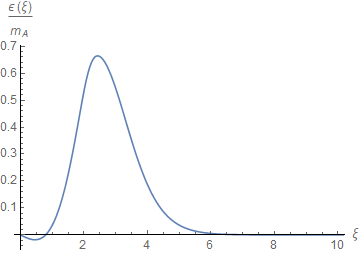}
\caption{Energy density $\epsilon (\xi) / \texttt{}m_A$ for $e=2$, $\lambda=1$, $\kappa=2$ ($c_1=1$), $\beta=160$, $\gamma=200$, $\mu / \phi_0=0.8$ (dimensionful parameters have been normalized by the gauge field mass).}
\label{fig:energiaregion1}
\end{figure}

The energy density of the vortex solution as a function of the dimensionless radial coordinate $\xi$ is shown in Fig. \ref{fig:energiaregion1}. It is important to notice  that at short distances the energy density is negative due to the fact that the potential $U[\vec{\chi},\vec{\phi}]$ is negative there and all other field contributions tend to zero when one approaches the origin.
 {Because of this fact, when  both orders are in competition, the total energy ${\cal E}_{CSH\chi}$ is  lower than the energy ${\cal E}_{CSH}$ of the pure CS-Higgs vortices. Indeed, in the latter case the BPS bound leads, for the minimal energy solution with topological charge  $n=1$ \cite{CLS} so that for the gauge coupling constant $e =2$  chosen in Fig. \ref{fig:energiaregion1}  the result is
\be
{\cal E}_{CSH} =\frac{e \phi_0^2}2
\pi = \pi \phi_0^2
\ee
while for the solution in which the competing $\chi$ field is present we get
\be
{\cal E}_{CSH\chi}
 =2.66623 \phi_0^2
 \ee
Hence the presence of the competing field $\chi$ lowers the energy by $ \Delta {\cal E} =- 0.778 \phi_0^2$.
This phenomenon has been reported in many studies of the so called intertwined orders in high temperature
superconductors (see \cite{EdRMP} and references there) and more recently in a variational study of a phenomenological nonlinear
sigma-model with two competing orders \cite{Edcom}.  }

\begin{figure}[h]
\centering
\includegraphics[scale=.5]{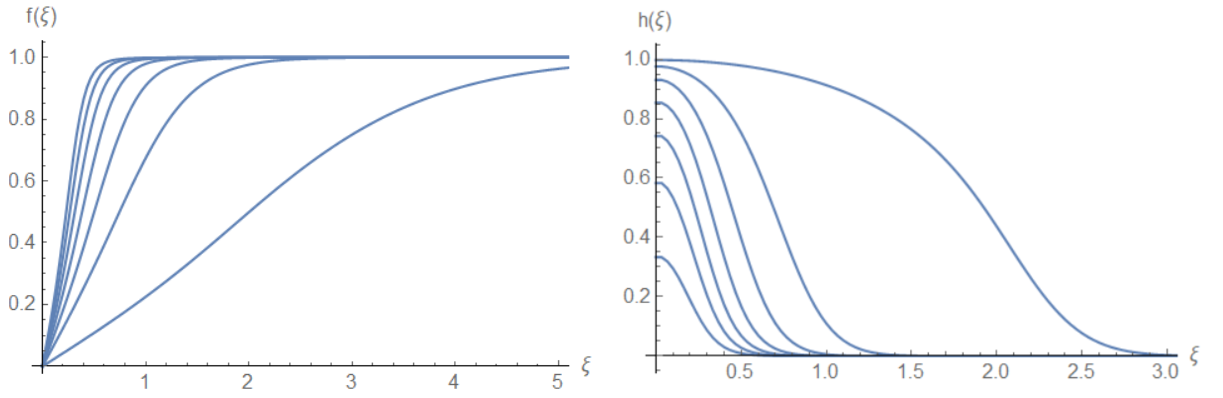}
\caption{Field profiles at distinct values of $m_\phi$. The figure on the left shows Higgs profiles $f(\xi)$, while in the right  one it is the competing $\chi$ field  profiles $h(\xi)$ that are shown. From right to left, the values of $\lambda$ grow according to $\lambda = 1,3,5,7,9,11,13$ so that the Higgs mass also grows consistently as $m_\phi = 2 \lambda \phi_0^2$. The other parameters remain fixed  at $e=2$, $\kappa=2$, $\beta=160$, $\gamma=200$, $\mu / \phi_0=0.8$ (dimensionful parameters have been normalized with respect to the gauge field mass). As it was to be expected,  as the mass grows larger, the Higgs field reaches earlier its vacuum value.}
\label{fig:multis}
\end{figure}

We have explored a wide range of parameter values. Fig. \ref{fig:multis} shows the effect of changing the Higgs mass $m_\phi=2\lambda \phi_0^2$ by changing $\lambda$ with all other parameters fixed.

As the Higgs mass grows, the size of the vortex core decreases, as can be seen in Fig. \ref{fig:multis} and  also the $\chi$ field profile shrinks both in height and extension. We find that the same happens  when one takes smaller $\beta$ and $\gamma$ values in such a way that   condition \eqref{escalaescalares} and stability requirements hold.

\begin{figure}[h]
\centering
\includegraphics[scale=.5]{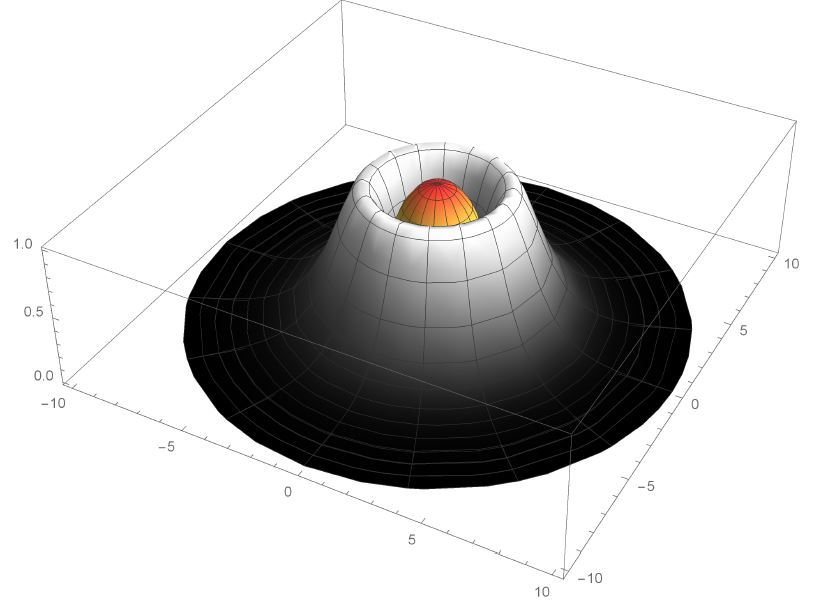}
\caption{In the  vertical axis we represent the $\chi$ field as a function of ($x,y$) in a color gradient from yellow to red, and around it the magnetic field modulus in a color gradient from white to black for $e=2$, $\lambda=1$, $\kappa=2$ ($c_1=1$), $\beta=160$, $\gamma=200$, $\mu / \phi_0=0.8$ (dimensionful parameters have been normalized by the gauge field mass).}
\label{fig:halo3d}
\end{figure}

In Fig. \ref{fig:halo3d} we have plotted the magnetic field and the $\chi$ field as a function of ($x,y$). Note that in contrast with what happens for  the Maxwell gauge dynamics, here  the magnetic field does vanish at the origin so that the core of the vortex is not a disk but an annulus. The $\chi$ halo therefore was found to be inside the smaller radius circle, as was mentioned before.

The   values in   figures 1 to 5 were chosen because they lead to an enhanced value of $\chi$. But in fact the region of competing orders was found for a wide range of parameter values.

\begin{figure}[H]
\centering
\includegraphics[scale=.5]{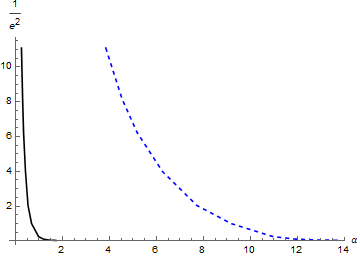}
\caption{Analysis of the solutions in different parameter regions (at $\lambda=1$, $\kappa=2$, $\beta=400$, $\gamma=200$, normalized with respect to the gauge mass).  Horizontal and vertical axis are $\alpha = \phi^2_0 / \mu^2$ and $1/e^2$ respectively. To the left of the solid line no vortex solutions were found. To the right of the dashed line there is no competing order and ordinary vortex CSH solutions can be obtained. In between these regions one finds vortex solutions with non-trivial $\chi$ order.}
\label{fig:curvasecuadrado}
\end{figure}

Based on our  numerical analysis of the solutions for different parameter ranges we present in
Fig. \ref{fig:curvasecuadrado}  the three regions in which the competing scalar field solutions are trivial, purely CS vortex-like or they show a competition between two order parameters. We have chosen for the horizontal axis $\alpha = \phi_0^2/\mu^2$ and  studied the fate of the solutions when one changes the value of the $\vec A_\mu-\vec\phi$ coupling constant $e$. We chose $1/e^2$ for the vertical axis because as it grows, the magnetic field contribution to the energy density also grows (see eq.\eqref{inserting}) so that the figure can be identified with the usual magnetic field versus control parameter diagrams.

In the region to the right of the dashed curve vortices are those of a CS-Higgs model as discussed in \cite{CLS}: no competing order with the $\chi$ field takes place. To the left of the solid line no non-trivial $(\phi,\chi)$ vortex solutions exist. Solely in the region between the solid and dashed lines there is a competing order where the $\phi$ and $\chi$ fields are non-trivial and behave as in Fig. \ref{figscalars:c1=1}.
 {As $\alpha$ grows, the ``halo'' produced by the $\chi$ field about the vortex core reduces in size until no competing order is left. This is consistent with the definition of $\alpha$ as $\phi_0^2/\mu^2$, since a big value of $\alpha$ would imply the Higgs field dominating over the competing $\chi$ field.}

Fig. \ref{fig:curvasecuadrado} can be interpreted at the light of  the schematic  zero temperature phase diagram of a layered superconductor in the presence of two order parameter proposed in reference \cite{Ed1}. To this end one can  identify the two axes in their diagram, the magnetic induction $B$ and a control parameter $\alpha$, with $1/e^2$ and our $\alpha$ parameter (note that when $1/e^2$ grows the contribution of the magnetic field to the energy density also grows).

 The  crossover lines in  the  phase diagram of ref. \cite{Ed1} can  be identified with the lines   separating the three regions in figure \ref{fig:curvasecuadrado} that we found  without any approximation from the  numerical solution of the coupled system of equations. It is also important to note that $\alpha_c=1$, the point  at which the constant non-trivial solution for the competing scalar vanishes corresponds in ref. \cite{Ed1} to the point  in which there is  a continuous
transition at $B=0$  between a pure superconducting phase and a
phase with a coexisting orders. In contrast, because   our potential
$U[\phi,\chi]$  has a larger number of parameters to fix,   suitable choices  allow to find solutions with coexisting orders both for $\alpha \lessgtr \alpha_c$, provided one is not  too far from $\alpha_c$.

The figure also shows why equation \eqref{escalaescalares} needs to be satisfied in order to have vortex solutions with a competing field halo. As $\alpha$ grows the halo reduces in size until the condition is no longer satisfied and we are left with ordinary CS vortices to the right of the dashed line. Physically this means that the Higgs field is dominant with respect to the competing field and the latter becomes irrelevant. A similar effect occurs when $\lambda$ grows and the RHS of condition \eqref{escalaescalares} becomes larger, as can be seen from figure \ref{fig:multis}. Finally, we observed the same behavior when changing the value of $\gamma$, considering that a large $\gamma$ implies a strong interaction between the fields and for small $\gamma$ the competing orders decouple and again, no halo can be found.

Moving to smaller $\alpha$, condition \eqref{escalaescalares} is easily satisfied and one would expect to still find vortex solutions with a competing halo. However, as mentioned before, there is another solution to the field equations playing a role in this region. This is the constant fields solution from eq. \eqref{solconstante}, valid for $\alpha < 1$. This solution has a lower energy  than the vortex configuration so it destabilizes the latter when one works with small values of $\alpha$. This argument provides a physical explanation to why numerically no vortex solutions were found at small values of the parameter $\alpha$. Note that topological stability does not in general guarantee that the solution cannot decay to a lower energy one. It happens, for example, in the case of the $O(3)$ sigma model as a size instability. This also takes place in the decay of vortex solutions. Indeed, when   the Landau parameter value is $\lambda > 1$, the energy of a vortex with topological charge $n=2$   is larger than that of two vortices with $n=1$, as it has been shown numerically \cite{Manton}.

\section{Summary and discussion}

In this work we have analyzed how the vortex-like solutions of an $SU(2)$ Chern-Simons-Higgs theory are affected when an uncharged scalar triplet $\chi$  is coupled to one of the Higgs scalars with a potential chosen so that $\chi$ develops a non-trivial vacuum state in the vortex core. As originally discussed in \cite{Shifman} this is a simple mechanism through which  vortex strings can acquire non-Abelian moduli localized on their world sheets.

Models with two competing vacuum states are also of interest in studies of superconductivity phenomena and in this respect it is interesting to compare our results with the analysis   of a layered superconductor in the vicinity of
a quantum critical point separating a pure superconducting phase from a phase in which a competing order
coexists with superconductivity \cite{Ed1}. The  main results in that paper  are summarized in the schematic zero temperature phase diagram that the authors propose which can be compared with the diagram presented in our Fig. \ref{fig:curvasecuadrado}.

Theories  where two sectors with different field content are coupled through a potential like the one we employed here are also actively investigated in connection with cosmological problems like that of the fractional cosmic neutrinos \cite{SW} and the dark matter problem \cite{Paola}. Their supersymmetric extension requires, apart from the (scalar-scalar) Higgs portal the addition of a kinetic gauge mixing \cite{AINS}. As a result, coupling constants accommodate to the Bogomolnyi point for which  the second order field equations can be reduced to the first  order ones originally found in the study of the Ginzburg-Landau theory of superconductivity when the Landau parameter takes the value $K =1/\sqrt2$  \cite{Harden}. We expect in future work to extend the analysis of the present work in some of the directions described above.
~

\subsection*{\bf Acknowledgements}
F.A.S. would like to thank Eduardo Fradkin for useful and estimulating discussions. The work of F.A.S. is supported by CONICET grant PIP 608  and FONCYT grant PICT 2304. The work by G.T. is supported by the Fondecyt grant 11160010.

\end{document}